\documentclass[journal, letter, onecolumn]{IEEEtran}
\usepackage{graphicx}
\usepackage{epsfig}
\usepackage{latexsym}
\usepackage{amsfonts}
\usepackage{here}
\usepackage{rawfonts}
\usepackage[latin1]{inputenc}
\usepackage[T1]{fontenc}
\usepackage{calc}
\usepackage{url}
\usepackage{enumerate}
\usepackage{color}
\usepackage[tbtags]{amsmath}
\usepackage{amssymb}
\usepackage{upref}
\usepackage{epic,eepic}
\usepackage{times}
\usepackage{dsfont}
\usepackage{comment}
\usepackage{cite}

\usepackage{graphicx}
\usepackage[font={small}]{caption}
\usepackage[font={small}]{subcaption}

\usepackage{stfloats}
\usepackage{mathtools}
\usepackage{amsmath}

\newtheorem{theorem}{\bf{Theorem}}
\newtheorem{definition}{\bf{Definition}}

\newtheorem{remark}{\bf{Remark}}

\ifCLASSINFOpdf
\else
\fi
\begin{document}

	\title{A Tractable Probability Distribution with Applications in Three-Dimensional Statistics}
	
	\author{{Seyed Mohammad Azimi-Abarghouyi}
		
		\thanks{S. M. Azimi-Abarghouyi is with the School of Electrical Engineering and Computer Science, KTH Royal Institute of Technology, Stockholm, Sweden (Email: seyaa@kth.se).}}
	
	\maketitle

\begin{abstract}
	This paper introduces and characterizes a new family of continuous probability distributions applicable to norm distributions in three-dimensional random spaces, specifically for the Euclidean norm of three random Gaussian variables with non-zero means. The distribution is specified over the semi-infinite range $[0,\infty)$ and is notable for its computational tractability. Building on this foundation, we also introduce a separate family of continuous probability distributions suitable for power distributions in three-dimensional random spaces. Despite being previously unknown, these distributions are attractive for numerous applications, some of which are discussed in this work.
\end{abstract}


\maketitle

\section{Introduction}
\label{sec:introduction}
To date, numerous probability distributions have been introduced and utilized in various branches of science, including the {\fontfamily{lmtt}\selectfont
	Rayleigh}, {\fontfamily{lmtt}\selectfont
	Rice}, and {\fontfamily{lmtt}\selectfont
	Nakagami} distributions, among others \cite{wikipedia, popolis,ross,nakagami,rice,ray}. For example, {\fontfamily{lmtt}\selectfont
	Rice} can model the distribution of Euclidean norms in two-dimensional random spaces. 

In this work, we introduce, define, and characterize a new family of probability distributions, including its special cases, which we name {\fontfamily{lmtt}\selectfont
	Abar} distribution\footnote{{\fontfamily{lmtt}\selectfont
		Abar} means superior in Persian language, and refers to applications of this distribution for higher dimensions than two.}. Also, we obtain its statistical measures in closed-forms. This distribution is recently discovered in our work \cite{azimi_molecular} for a research in biological communication networks. However, to the best of our knowledge, this distribution has not been previously discovered. For example, a similar distribution cannot be found in websites and books on "probability and statistics" \cite{wikipedia, popolis,ross}. We investigate  use-case applications for the {\fontfamily{lmtt}\selectfont
	Abar} distribution that are applicable to Euclidean norm or distance statistics in three-dimensional random spaces. Our focus is on revealing that
this distribution admits a remarkably simple closed form, which is not even the case in the simpler two-dimensional statistics. The proposed distribution has a unique and completely different function structure that is impossible to be extrapolated from its lower or higher dimensional cases, which all include integrals in their functions. Using the {\fontfamily{lmtt}\selectfont
	Abar} distribution, we also find another new family of probability distributions with its special cases, which we name {\fontfamily{lmtt}\selectfont
	Abar+} distribution. It is applicable to power statistics within three-dimensional random spaces. Due to their remarkable tractability and general applicability, the {\fontfamily{lmtt}\selectfont
	Abar} and {\fontfamily{lmtt}\selectfont
	Abar+} distributions hold promise for a wide range of applications, particularly in areas such as communication networks, localization and signal processing, stochastic geometry, and machine learning. By highlighting these distributions and the potential applications, researchers will become aware of their existence and have the opportunity to utilize them effectively.

\section{Distribution Characterization}
\begin{definition}
	Let $a$ and $\sigma$ two non-negative parameters. We define the {\fontfamily{lmtt}\selectfont
		Abar} distribution with the following probability density function (PDF)
	\begin{align}
	\label{abar}
	f_{a,\sigma}(y) &= \frac{y}{\sqrt{2\pi}\sigma a}\hspace{-3pt}\left[\exp\left(-\frac{(y-a)^2}{2\sigma^2}\right)-\exp\left(-\frac{(y+a)^2}{2\sigma^2}\right)\right] \nonumber\\
	&= \sqrt{\frac{2}{\pi}}\frac{y}{\sigma a} \exp\left(-\frac{y^2+a^2}{2\sigma^2}\right) \sinh \left(\frac{ya}{\sigma^2}\right); \ y\geq 0,
	\end{align}\footnote{Since $\exp\left(-\frac{(y-a)^2}{2\sigma^2}\right)\geq \exp\left(-\frac{(y+a)^2}{2\sigma^2}\right), \forall y$ in \eqref{abar}, $f_{a,\sigma}(y) \geq 0, \forall y$.}
	which is specialized for $a = 0$ as
	\begin{align}
	f_{0,\sigma}(y) &= \lim_{a \to 0} y \frac{ \frac{y-a}{\sigma^2}\exp\left(-\frac{(y-a)^2}{2\sigma^2}\right)+\frac{y+a}{\sigma^2}\exp\left(-\frac{(y+a)^2}{2\sigma^2}\right)}{\sqrt{2\pi}\sigma}  = \sqrt{\frac{2}{\pi}} \frac{y^2}{\sigma^3}\exp\left(-\frac{y^2}{2\sigma^2}\right); \ y\geq 0,
	\end{align}
	where the L'Hopital's rule is used.\footnote{The special case $a=0$ is also found in \cite{azimi_molecular}.} 
	
	Further, for $a \to \infty$, \eqref{abar} is specialized to 
	\begin{align}
	f_{\infty,\sigma}(y) &=\lim_{a \to \infty}\frac{\frac{y-a}{a}+1}{\sqrt{2\pi}\sigma}\exp\left(-\frac{(y-a)^2}{2\sigma^2}\right) = \frac{1}{\sqrt{2\pi}\sigma}\exp\left(-\frac{(y-a)^2}{2\sigma^2}\right); \ y\geq 0,
	\end{align}
	which is equal to the PDF of a Gaussian distribution with mean $a$ and variance $\sigma^2$, denoted as ${\cal N}(a, \sigma^2)$. 
	
	In Figs 1 and 2, the defined PDF in \eqref{abar} is shown for different parameters $a$ and $\sigma$. 
\end{definition}

\begin{remark}
		The structure of the {\fontfamily{lmtt}\selectfont
			Abar} distribution's PDF is fundamentally different from that of the Generalized Folded Normal ({\fontfamily{lmtt}\selectfont
			GFN}) distribution \cite{GFN}. This is primarily because the {\fontfamily{lmtt}\selectfont
			GFN} is characterized by the positive addition of two exponential functions, whose scaling factors are not dependent on the variable, contrasting with the structure of the {\fontfamily{lmtt}\selectfont
			Abar} distribution.
\end{remark}

From \eqref{abar}, the cumulative distribution function (CDF) of the {\fontfamily{lmtt}\selectfont
	Abar} distribution is obtained as
\begin{align}
\label{cdf}
&F_{a,\sigma}(y)  = 1- \frac{1}{\sqrt{2\pi}\sigma a}\int_{y}^{\infty}ye^{-\frac{(y-a)^2}{2\sigma^2}}-ye^{-\frac{(y+a)^2}{2\sigma^2}}\mathrm{d}y=1-\frac{\sigma}{\sqrt{2\pi} a}\int_{y}^{\infty}\frac{y-a}{\sigma^2}e^{-\frac{(y-a)^2}{2\sigma^2}}-\frac{y+a}{\sigma^2}e^{-\frac{(y+a)^2}{2\sigma^2}}\mathrm{d}y\nonumber\\&-\frac{1}{\sqrt{2\pi} \sigma}\int_{y}^{\infty}e^{-\frac{(y-a)^2}{2\sigma^2}}+e^{-\frac{(y+a)^2}{2\sigma^2}}\mathrm{d}y=1-\frac{\sigma}{\sqrt{2\pi} a}\left(e^{-\frac{(a-y)^2}{2\sigma^2}}-e^{-\frac{(a+y)^2}{2\sigma^2}}\right)-\frac{1}{2}\biggl(\frac{2}{\sqrt{\pi}} \int_{-\frac{a-y}{\sqrt{2}\sigma}}^{\infty} e^{-t^2}\mathrm{d}t+\nonumber\\&\frac{2}{\sqrt{\pi}} \int_{\frac{a+y}{\sqrt{2}\sigma}}^{\infty} e^{-t^2}\mathrm{d}t\biggr)=- \frac{\sigma}{\sqrt{2\pi} a} \left(e^{-\frac{(a-y)^2}{2\sigma^2}}-e^{-\frac{(a+y)^2}{2\sigma^2}}\right)+\frac{1}{2}\left(\text{erfc}\left(\frac{a-y}{\sqrt{2}\sigma}\right)-\text{erfc}\left(\frac{a+y}{\sqrt{2}\sigma}\right)\right),
\end{align}
where $\text{erfc}(x) =\frac{2}{\sqrt{\pi}}\int_{x}^{\infty}e^{-t^2}\mathrm{d}t$. It can be observed in \eqref{cdf} that $\frac{a}{\sigma}$ determines the ratio of the first part with exponential functions to the second part with erfc functions. Also, $F_{a,\sigma}(0) = 0$ and $F_{a,\sigma}(\infty) = 1$, which justifies the validity of the PDF in \eqref{abar}.

The mean of this distribution is
\begin{align}
\label{mean}
&\mathbb{E}_{a,\sigma}\left\{Y\right\} = \int_{0}^{\infty}\frac{y^2}{\sqrt{2\pi}\sigma a}\Biggl[\exp\left(-\frac{(y-a)^2}{2\sigma^2}\right)-\exp\left(-\frac{(y+a)^2}{2\sigma^2}\right)\Biggr]\mathrm{d}y= \int_{-\frac{a}{\sqrt{2}\sigma}}^{\infty} \frac{(\sqrt{2}\sigma t+a)^2}{\sqrt{\pi}a}e^{-t^2}\mathrm{d}t - \nonumber\\&\int_{\frac{a}{\sqrt{2}\sigma}}^{\infty} \frac{(\sqrt{2}\sigma t-a)^2}{\sqrt{\pi}a}e^{-t^2}\mathrm{d}t = \frac{2\sigma^2}{\sqrt{\pi}a} \Biggl[\frac{\sqrt{\pi}}{4}(1-\text{erfc}(t))-\frac{t}{2}e^{-t^2}\Biggr]_{-\frac{a}{\sqrt{2}\sigma}}^{\infty} + \frac{2\sqrt{2}\sigma}{\sqrt{\pi}} \left[-\frac{1}{2}e^{-t^2}\right]_{-\frac{a}{\sqrt{2}\sigma}}^{\infty} +\frac{a}{\sqrt{\pi}}\times\nonumber\\
&\left[\frac{\sqrt{\pi}}{2}\left(1-\text{erfc}(t)\right)\right]_{-\frac{a}{\sqrt{2}\sigma}}^{\infty}-\frac{2\sigma^2}{\sqrt{\pi}a} \Biggl[\frac{\sqrt{\pi}}{4}(1-\text{erfc}(t))-\frac{t}{2}e^{-t^2}\Biggr]_{\frac{a}{\sqrt{2}\sigma}}^{\infty} + \frac{2\sqrt{2}\sigma}{\sqrt{\pi}} \left[-\frac{1}{2}e^{-t^2}\right]_{\frac{a}{\sqrt{2}\sigma}}^{\infty} -\frac{a}{\sqrt{\pi}}\times\nonumber\\&\left[\frac{\sqrt{\pi}}{2}\left(1-\text{erfc}(t)\right)\right]_{\frac{a}{\sqrt{2}\sigma}}^{\infty} = \left(a+\frac{\sigma^2}{a}\right) \left(1-\text{erfc}\left(\frac{a}{\sqrt{2}\sigma}\right)\right)+\sqrt{\frac{2}{\pi}}\sigma \exp\left(-\frac{a^2}{2\sigma^2}\right).
\end{align}
It can be concluded from \eqref{mean}, when $\frac{a}{\sigma} \to 0$, we have 
\begin{align}
&\mathbb{E}_{a,\sigma}\left\{Y\right\} \to \left(a+\frac{\sigma^2}{a}\right) \left(1-\text{erfc}\left(\frac{a}{\sqrt{2}\sigma}\right)\right)+\sqrt{\frac{2}{\pi}}\sigma \to  \frac{1-\text{erfc}\left(\frac{a}{\sqrt{2}\sigma}\right)}{\frac{\frac{a}{\sigma^2}}{\frac{a^2}{\sigma^2}+1}}+\sqrt{\frac{2}{\pi}}\sigma \to  \frac{1-\text{erfc}\left(\frac{a}{\sqrt{2}\sigma}\right)}{{\frac{a}{\sigma^2}}}+\sqrt{\frac{2}{\pi}}\sigma \stackrel{(a)}{\to}\nonumber\\& \frac{-\frac{2}{\sqrt{\pi}} \times -\frac{1}{\sqrt{2}}\exp\left(-\frac{a^2}{2\sigma^2}\right)}{\frac{1}{\sigma}}+\sqrt{\frac{2}{\pi}}\sigma \to \sqrt{\frac{2}{\pi}}\sigma + \sqrt{\frac{2}{\pi}}\sigma = 2 \sqrt{\frac{2}{\pi}}\sigma,
\end{align}
where $(a)$ comes from the L'Hopital's rule. This result denotes the fact that the terms $\sqrt{\frac{2}{\pi}}\sigma \exp\left(-\frac{a^2}{2\sigma^2}\right)$ and \\$\left(a+\frac{\sigma^2}{a}\right) \left(1-\text{erfc}\left(\frac{a}{\sqrt{2}\sigma}\right)\right)$ in \eqref{mean} contribute the same mean when $\frac{a}{\sigma} \to 0$. Also, when $\frac{a}{\sigma} \to \infty$, we have $\mathbb{E}_{a,\sigma}\left\{Y\right\} \to a+\frac{\sigma^2}{a} = a \left(1+\frac{\sigma^2}{a^2}\right) \to a$. 

For the variance, first 
\begin{align}
&\mathbb{E}_{a,\sigma}\left\{Y^2\right\} = \int_{0}^{\infty}\frac{y^3}{\sqrt{2\pi}\sigma a}\Biggl[\exp\left(-\frac{(y-a)^2}{2\sigma^2}\right)-\exp\left(-\frac{(y+a)^2}{2\sigma^2}\right)\Biggr]\mathrm{d}y= \int_{-\frac{a}{\sqrt{2}\sigma}}^{\infty} \frac{(\sqrt{2}\sigma t+a)^3}{\sqrt{\pi}a}e^{-t^2}\mathrm{d}t -\nonumber\\
& \int_{\frac{a}{\sqrt{2}\sigma}}^{\infty} \frac{(\sqrt{2}\sigma t-a)^3}{\sqrt{\pi}a}e^{-t^2}\mathrm{d}t =\frac{2\sqrt{2}\sigma^3}{\sqrt{\pi}a}\left[-\frac{e^{-t^2}}{2}\left(t^2+1\right)\right]_{-\frac{a}{\sqrt{2}\sigma}}^{\infty}+\frac{6\sigma^2}{\sqrt{\pi}}\left[\frac{\sqrt{\pi}}{4}(1-\text{erfc}(t))-\frac{t}{2}e^{-t^2}\right]_{-\frac{a}{\sqrt{2}\sigma}}^{\infty} +\frac{3\sqrt{2}\sigma a}{\sqrt{\pi}}\times\nonumber\\
&\left[-\frac{1}{2}e^{-t^2}\right]_{-\frac{a}{\sqrt{2}\sigma}}^{\infty}+\frac{a^2}{\sqrt{\pi}}\left[\frac{\sqrt{\pi}}{2}\left(1-\text{erfc}(t)\right)\right]_{-\frac{a}{\sqrt{2}\sigma}}^{\infty}-\frac{2\sqrt{2}\sigma^3}{\sqrt{\pi}a}\left[-\frac{e^{-t^2}}{2}\left(t^2+1\right)\right]_{\frac{a}{\sqrt{2}\sigma}}^{\infty}+\frac{6\sigma^2}{\sqrt{\pi}}\Biggl[\frac{\sqrt{\pi}}{4}(1-\text{erfc}(t))-\nonumber
\end{align}
\begin{align}
&\frac{t}{2}e^{-t^2}\Biggr]_{\frac{a}{\sqrt{2}\sigma}}^{\infty} -\frac{3\sqrt{2}\sigma a}{\sqrt{\pi}}\left[-\frac{1}{2}e^{-t^2}\right]_{\frac{a}{\sqrt{2}\sigma}}^{\infty}+\frac{a^2}{\sqrt{\pi}}\left[\frac{\sqrt{\pi}}{2}\left(1-\text{erfc}(t)\right)\right]_{\frac{a}{\sqrt{2}\sigma}}^{\infty} = 3 \sigma^2 + a^2,
\end{align}
which is a very simple linear function of the parameters of the PDF, $a$ and $\sigma$, and then
\begin{align}
&\mathbb{V}\text{ar}_{a,\sigma}\left\{Y\right\} = \mathbb{E}_{a,\sigma}\left\{Y^2\right\} - \mathbb{E}_{a,\sigma}^2\left\{Y\right\} =3 \sigma^2 + a^2 -\left(a+\frac{\sigma^2}{a}\right)^2 \left(1-\text{erfc}\left(\frac{a}{\sqrt{2}\sigma}\right)\right)^2-{\frac{2}{\pi}}\sigma^2 \exp\left(-\frac{a^2}{\sigma^2}\right)\nonumber\\&- 2\sqrt{\frac{2}{\pi}}\sigma\left(a+\frac{\sigma^2}{a}\right) \left(1-\text{erfc}\left(\frac{a}{\sqrt{2}\sigma}\right)\right) \exp\left(-\frac{a^2}{2\sigma^2}\right).
\end{align}
When $\frac{a}{\sigma} \to 0$, we have $\mathbb{V}\text{ar}_{a,\sigma}\left\{Y\right\} \to (3-{\frac{8}{\pi}}) \sigma^2 + a^2 = \sigma^2 \left((3-{\frac{8}{\pi}})  + \frac{a^2}{\sigma^2}\right) = (3-{\frac{8}{\pi}})\sigma^2$. Also, when $\frac{a}{\sigma} \to \infty$, $\mathbb{V}\text{ar}_{a,\sigma}\left\{Y\right\} \to 3\sigma^2+a^2 - (a+\frac{\sigma^2}{a})^2 \to \sigma^2\left(1-\frac{\sigma^2}{a^2}\right) \to \sigma^2$. 

Finally, the moment generating function (MGF) of this distribution is obtained as
\begin{align}
&M_{a,\sigma}(s) =\mathbb{E}\left\{e^{sY}\right\}= \int_{0}^{\infty} \frac{y}{\sqrt{2\pi}\sigma a}\Biggl[\exp\left(sy-\frac{(y-a)^2}{2\sigma^2}\right)-\exp\left(sy-\frac{(y+a)^2}{2\sigma^2}\right)\Biggr]\mathrm{d}y = \int_{0}^{\infty} \frac{y}{\sqrt{2\pi}\sigma a}\times\nonumber\\&\Biggl[\exp\left(-\frac{1}{2\sigma^2}\left(y-(a+s\sigma^2)\right)^2+s\left(a+\frac{s\sigma^2}{2}\right)\right)-\exp\left(-\frac{1}{2\sigma^2}\left(y+(a-s\sigma^2)\right)^2-s\left(a-\frac{s\sigma^2}{2}\right)\right)\Biggr]\mathrm{d}y = \nonumber\\
&e^{s\left(a+\frac{s\sigma^2}{2}\right)} \int_{-\frac{a+s\sigma^2}{\sqrt{2}\sigma}}^{\infty} \frac{\sqrt{2}\sigma t+(a+s\sigma^2)}{\sqrt{\pi}a}e^{-t^2}\mathrm{d}t-e^{-s\left(a-\frac{s\sigma^2}{2}\right)} \int_{\frac{a-s\sigma^2}{\sqrt{2}\sigma}}^{\infty} \frac{\sqrt{2}\sigma t-(a-s\sigma^2)}{\sqrt{\pi}a}e^{-t^2}\mathrm{d}t = e^{s\left(a+\frac{s\sigma^2}{2}\right)}\Biggl[-\frac{\sigma}{\sqrt{2\pi}a}\times\nonumber\\&e^{-t^2}\bigg|_{-\frac{a+s\sigma^2}{\sqrt{2}\sigma}}^{\infty}+\frac{a+s\sigma^2}{2a}\frac{2}{\sqrt{\pi}}\int_{-\frac{a+s\sigma^2}{\sqrt{2}\sigma}}^{\infty}e^{-t^2}\mathrm{d}t\Biggr]-e^{-s\left(a-\frac{s\sigma^2}{2}\right)}\left[-\frac{\sigma}{\sqrt{2\pi}a}e^{-t^2}\bigg|_{\frac{a-s\sigma^2}{\sqrt{2}\sigma}}^{\infty}-\frac{a-s\sigma^2}{2a}\frac{2}{\sqrt{\pi}}\int_{\frac{a-s\sigma^2}{\sqrt{2}\sigma}}^{\infty}e^{-t^2}\mathrm{d}t\right]\nonumber\\
&=e^{s\left(a+\frac{s\sigma^2}{2}\right)}\Biggl[\frac{\sigma}{\sqrt{2\pi}a}e^{-\frac{(a+s\sigma^2)^2}{2\sigma^2}}+\frac{a+s\sigma^2}{2a}\left(2-\text{erfc}\left(\frac{a+s\sigma^2}{\sqrt{2}\sigma}\right)\right)\Biggr]-e^{-s\left(a-\frac{s\sigma^2}{2}\right)}\Biggl[\frac{\sigma}{\sqrt{2\pi}a}e^{-\frac{(a-s\sigma^2)^2}{2\sigma^2}}\nonumber\\
&-\frac{a-s\sigma^2}{2a}\text{erfc}\left(\frac{a-s\sigma^2}{\sqrt{2}\sigma}\right)\Biggr].
\end{align}
It is notable that all the results in (1)-(9) are in tractable closed-forms. This fact renders the {\fontfamily{lmtt}\selectfont
	Abar} distribution extremely attractive for use in various applications.
\vspace{0pt}
\section{Applications}
We have discovered two applications that exhibit the {\fontfamily{lmtt}\selectfont
	Abar} distribution, specifically relating to the norm of three Gaussian random variables. In contrast, applications involving the norm of two Gaussian random variables result in the {\fontfamily{lmtt}\selectfont
	Rice} distribution \cite{rice, afshang}. Unlike the {\fontfamily{lmtt}\selectfont
	Abar} distribution, the {\fontfamily{lmtt}\selectfont
	Rice} distribution cannot be expressed in a closed form and requires numerical computations involving Bessel functions. Additionally, it possesses a distinct structure compared to the {\fontfamily{lmtt}\selectfont
	Abar} distribution. The simplicity found in the structure of the {\fontfamily{lmtt}\selectfont
	Abar} distribution is fundamentally dependent on the three-dimensional assumption and cannot be inferred from cases with lower or higher dimensions. The unique and independent nature of the {\fontfamily{lmtt}\selectfont Abar} distribution, even in comparison with the {\fontfamily{lmtt}\selectfont Rice} distribution, raises an intriguing question: how can a simpler, closed-form function for the three-dimensional case emerge from the more complex, integral-based function characteristic of the two-dimensional case. This question is addressed in the following subsection. In the third application, we employ the {\fontfamily{lmtt}\selectfont
	Abar} distribution to derive yet another new tractable distribution relevant to the squared norm of three Gaussian random variables.
\begin{figure}[t!]
	\vspace{0pt}
	\centering
	\includegraphics[width =3.2in]{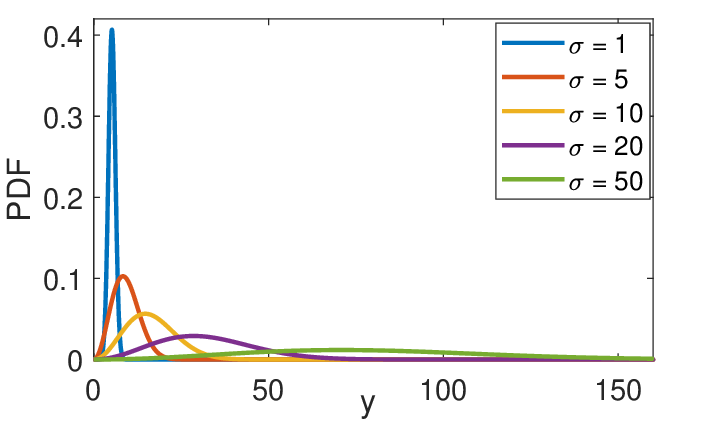} 
	\vspace{0pt}
	\caption{The {\fontfamily{lmtt}\selectfont
			Abar} PDF when $a = 5$.}
	\vspace{-5pt}
\end{figure}

\begin{figure}[t!]
	\vspace{0pt}
	\centering
	\includegraphics[width =3.2in]{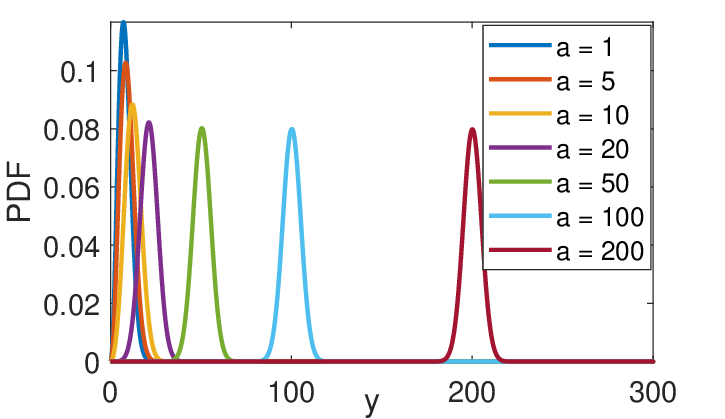} 
	\vspace{0pt}
	\caption{The {\fontfamily{lmtt}\selectfont
			Abar} PDF when $\sigma = 5$.}
	\vspace{-5pt}
\end{figure}

\subsection{Application 1} It is presented in the following theorem. \begin{theorem} 
	Assume the norm $y = \sqrt{y_1^2+y_2^2+y_3^2}$, where $y_1$, $y_2$, and $y_3$ are independent Gaussian random variables with the same variance $\sigma^2$ and means $a_1$, $a_2$, and $a_3$, respectively. Also, let $a = \sqrt{a_1^2+a_2^2+a_3^2}$. Then, $y$ has {\fontfamily{lmtt}\selectfont
		Abar} distribution with parameters $a$ and $\sigma$. 
\end{theorem}
\begin{IEEEproof} We can write $y_i = z_i+a_i, i \in \left\{1,2,3\right\}$ where $z_i, i\in \left\{1,2,3\right\}$ are independent identical Gaussian random variables with mean zero and variance $\sigma^2$. Then, we have
	\begin{align}
	\label{pdfproof}
	&\mathbb{P}\left(\sqrt{y_1^2+y_2^2+y_3^2}<y\right)= \int_{z_1=-y}^{y}\int_{z_2 = -\sqrt{y^2-z_1^2}}^{\sqrt{y^2-z_1^2}}\int_{z_3=-\sqrt{y^2-z_1^2-z_2^2}}^{\sqrt{y^2-z_1^2-z_2^2}}\frac{1}{(2\pi)^\frac{3}{2} \sigma^3}\times\nonumber\\
	&\exp\left(-\frac{(z_1-a_1)^2+(z_2-a_2)^2+(z_3-a_3)^2}{2 \sigma^2}\right)\mathrm{d}z_3\mathrm{d}z_2\mathrm{d}z_1\nonumber\\
	&=\int_{z_1=-y}^{y}\int_{z_2 = -\sqrt{y^2-z_1^2}}^{\sqrt{y^2-z_1^2}}\int_{z_3=-\sqrt{y^2-z_1^2-z_2^2}}^{\sqrt{y^2-z_1^2-z_2^2}}\frac{1}{(2\pi)^\frac{3}{2} \sigma^3}\exp\left(-\frac{(z_1-a)^2+z_2^2+z_3^2}{2 \sigma^2}\right)\mathrm{d}z_3\mathrm{d}z_2\mathrm{d}z_1,
	\end{align}
	where the last equality is because Gaussian random variables are rotationally invariant. Then, by taking derivative of \eqref{pdfproof}, the PDF of $y$ is obtained with the help of the Leibniz integral rule and then
	\begin{align}
	&f_{a,\sigma}(y) =  \frac{2y}{(2\pi)^\frac{3}{2}\sigma^3}\exp\left(-\frac{y^2+a^2}{2\sigma^2}\right) \int_{z_1=-y}^{y}\exp\left(\frac{az_1}{\sigma^2}\right)\int_{z_2 = -\sqrt{y^2-z_1^2}}^{\sqrt{y^2-z_1^2}}\frac{1}{\sqrt{y^2-z_1^2-z_2^2}}\mathrm{d}z_2\mathrm{d}z_1,
	\end{align}
	where 
	\begin{align}
	\label{reason}
	\int_{z_2 = -\sqrt{y^2-z_1^2}}^{\sqrt{y^2-z_1^2}}\frac{1}{\sqrt{y^2-z_1^2-z_2^2}}\mathrm{d}z_2 = \pi.
	\end{align}
	The constant equality in \eqref{reason} plays the crucial role for providing a simple closed-form PDF structure, which is not the possible case in lower or higher dimensions (since integrals persist in the final results). 
	
	Finally, we obtain
	\begin{align}
	f_{a,\sigma}(y) &= \frac{y}{\sqrt{2\pi}\sigma a}\exp\left(-\frac{y^2+a^2}{2\sigma^2}\right)\left[\exp\left(\frac{ya}{\sigma^2}\right)-\exp\left(-\frac{ya}{\sigma^2}\right)\right]=\nonumber\\&
	\frac{y}{\sqrt{2\pi}\sigma a}\left[\exp\left(-\frac{(y-a)^2}{2\sigma^2}\right)-\exp\left(-\frac{(y+a)^2}{2\sigma^2}\right)\right].
	\end{align}

\end{IEEEproof}

\subsection{Application 2} It is shown in our work \cite[Th.1]{azimi_molecular} that the distance distributions in the three-dimensional Thomas cluster process (TCP) have the {\fontfamily{lmtt}\selectfont
	Abar} distribution.\footnote{Application 1 is the abstract form of Application 2.} This pertains to the distance from each cluster center to any point in other clusters. TCP is a widely-used class of Poisson cluster processes in the field of stochastic geometry and has numerous applications in various domains such as biology and communication networks, serving as a valuable tool for modeling and analysis purposes \cite{afshang, azimi_molecular, haenggi_book, azimi_cluster1}. Given that the spatial domain in these fields is typically represented by the three dimensions $(x,y,z)$, the PDF introduced and the findings presented in this work are anticipated to be of significant value for evaluating spatial patterns and conducting performance analyses. For a practical and real-time numerical example, please see \cite[Sections IV, V, and VI]{azimi_molecular} for the utilization and crucial application of {\fontfamily{lmtt}\selectfont
		Abar} distribution in deriving the expected interference value and the error probability in communication nanonetworks, where nanomachine sensors are located based on three-dimensional TCP. An alternative application worth considering is the network of clustered unmanned aerial vehicles (UAVs), where three-dimensional TCP can be utilized for modeling the locations of UAVs in the sky. 

In addition, the distance distribution is a widely used metric in signal processing and communication systems. For example, it can be used in target localization, such as satellite localization or range detection in radar systems, and also satellite communication systems. In these applications, there are three location dimensions and two-dimensional modeling is not accurate. Also, it has applications in signal (modulation) classification and outlier detection, where distances of signals from each other and the noise are important features.

\begin{figure}[t!]
	\vspace{0pt}
	\centering
	\includegraphics[width =3.2in]{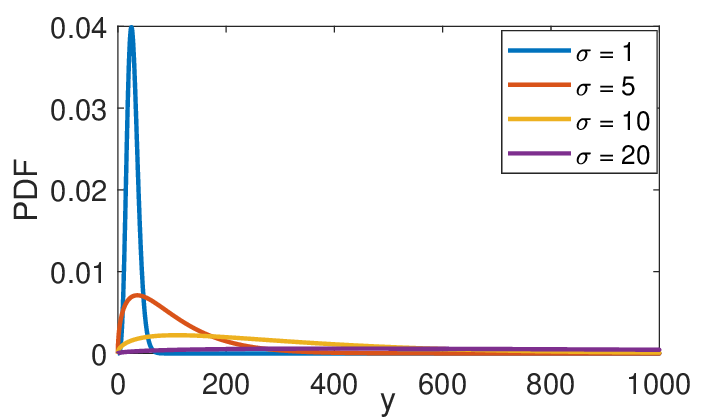} 
	\vspace{0pt}
	\caption{The {\fontfamily{lmtt}\selectfont
			Abar+} PDF when $a = 5$.}
	\vspace{-5pt}
\end{figure}

\begin{figure}[t!]
	\vspace{0pt}
	\centering
	\includegraphics[width =3.2in]{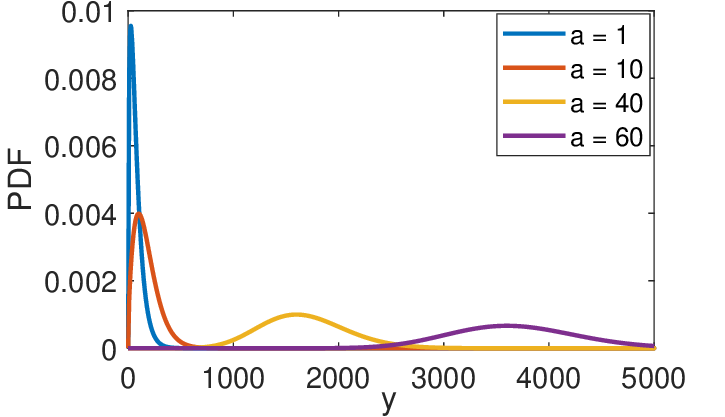} 
	\vspace{0pt}
	\caption{The {\fontfamily{lmtt}\selectfont
			Abar+} PDF when $\sigma = 5$.}
	\vspace{-5pt}
\end{figure}

\subsection{Application 3} The following theorem presents it.
\begin{theorem}
	Assume $y = {y_1^2+y_2^2+y_3^2}$, where $y_1$, $y_2$, and $y_3$ are independent Gaussian random variables with the same variance $\sigma^2$ and means $a_1$, $a_2$, and $a_3$, respectively. Also, let $a = \sqrt{a_1^2+a_2^2+a_3^2}$. Then, $y$ has the following PDF with parameters $a$ and $\sigma$
	\begin{align}
	\label{abar+}
	f_{a,\sigma}^{+}(y) &= \frac{1}{2\sqrt{2\pi} a\sigma} \Biggl[ \exp\left({-\frac{(a-\sqrt{y})^2}{2\sigma^2}}\right) - \exp\left({-\frac{(a+\sqrt{y})^2}{2\sigma^2}}\right) \Biggr]\nonumber\\&= \frac{1}{\sqrt{2\pi}a \sigma} \exp \left(-\frac{a^2+y}{2\sigma^2}\right) \sinh \left(\frac{a\sqrt{y}}{\sigma^2}\right); \ y \geq 0.
	\end{align}
\end{theorem}
\begin{IEEEproof}
	The CDF of $y$ can be obtained from the CDF of {\fontfamily{lmtt}\selectfont
		Abar} as
	\begin{align}
	\label{abar+cdf}
	F_{a,\sigma}^+(y) & =\mathbb{P}(Y<y) = \mathbb{P}(\sqrt{Y}<\sqrt{y}) = F_{a,\sigma}(\sqrt{y}) \nonumber\\&= - \frac{\sigma}{\sqrt{2\pi} a} \left(e^{-\frac{(a-\sqrt{y})^2}{2\sigma^2}}-e^{-\frac{(a+\sqrt{y})^2}{2\sigma^2}}\right)+\frac{1}{2}\left(\text{erfc}\left(\frac{a-\sqrt{y}}{\sqrt{2}\sigma}\right)-\text{erfc}\left(\frac{a+\sqrt{y}}{\sqrt{2}\sigma}\right)\right),
	\end{align}
	where $F_{a,\sigma}({y})$ is given in \eqref{cdf}. Then, by taking derivative of \eqref{abar+cdf} and using $\frac{\mathrm{d}}{\mathrm{d}x}\text{erfc}(x) = -\frac{2}{\sqrt{\pi}} e^{-x^2}$, the proof is complete.		
\end{IEEEproof}
As the PDF in \eqref{abar+} has not been previously discovered \cite{wikipedia, popolis,ross}, we define its distribution with the name {\fontfamily{lmtt}\selectfont
	Abar+}.\footnote{This distribution, not present in \cite{azimi_molecular}, is discovered in the current study.} The parameters $a$ and $\sigma$ are non-negative. In Figs 3 and 4, the defined PDF in \eqref{abar+} is shown for different parameters $a$ and $\sigma$. Using the L'Hopital's rule, the special case of this distribution for $a = 0$ is given for $y \geq 0$ as
\begin{align}
&f_{0,\sigma}^{+}(y) = \lim_{a \to 0}\frac{-\frac{1}{\sigma^2} (a-\sqrt{y}) \exp\left({-\frac{(a-\sqrt{y})^2}{2\sigma^2}}\right)}{2\sqrt{2\pi}\sigma}+  \frac{\frac{1}{\sigma^2} (a+\sqrt{y}) \exp\left({-\frac{(a+\sqrt{y})^2}{2\sigma^2}}\right)}{2\sqrt{2\pi}\sigma}=\frac{\sqrt{y}}{\sqrt{2\pi}\sigma^3} \exp\left(-\frac{y}{2\sigma^2}\right),
\end{align}
which is equal to the Gamma distribution with the shape $\frac{3}{2}$ and scale ${2\sigma^2}$, denoted as ${\fontfamily{lmtt}\selectfont
	Gamma}(\frac{3}{2},\frac{1}{2\sigma^2})$. In addition, for $a \to \infty$, \eqref{abar+} goes to the all-zero function. 

The {\fontfamily{lmtt}\selectfont
		Abar+} distribution has the following statistical measures.\footnote{The proof has not been included.}

Mean:
	\begin{align}
	&\mathbb{E}_{a,\sigma}^+\left\{Y\right\} = a^2 + 3 \sigma^2.
	\end{align}

Variance:
	\begin{align}
	&\mathbb{V}\text{ar}_{a,\sigma}^+\left\{Y\right\} = 4 a^2 \sigma^2 + 6 \sigma^4.
	\end{align}

MGF:
	\begin{align}
	&M_{a,\sigma}^+(s) = \frac{1}{(1 - 2 s \sigma^2)^{\frac{3}{2}}}\exp\left(\frac{a^2 s}{1 - 2 s \sigma^2}\right).
	\end{align}

\begin{remark}
		The Gamma distribution is associated with the sum of squared independent Gaussian random variables that have a zero mean. In contrast, the {\fontfamily{lmtt}\selectfont Abar+} distribution incorporates Gaussian variables with distinct, non-zero means.
\end{remark}
As signal samples can be well approximately modeled by Gaussian random variables, where the mean represents the bias of signal, Application 3 possesses exceptional worth in the modeling and analysis of the power of additive signals, as well as in their subsequent processing. In wireless multipath environments, for instance, there exist three main signal paths between a transmitter and a receiver, which can be represented as Gaussian random variables with different means: 1) the direct link, 2) the reflection from ground nodes and surfaces, and 3) the reflection from sky nodes, such as UAVs and satellites. 

It is our belief that there exists a wealth of additional and potentially unknown applications within the fields of pure and applied mathematics and other scientific domains, specially communications and signal processing. Furthermore, due to the remarkable tractability and closed-form structures, the outcomes of this work hold significant potential for applications in the areas of statistics and machine learning, particularly with regards to classification and clustering, where norm distributions can be leveraged \cite{wanright, murphy}. To be more specific, clustering is a technique that groups similar data points together based on their distance from each other. By modeling the distances between data points, clustering algorithms can be optimized to better identify and separate distinct clusters within the data. It is worth mentioning that higher dimensional spaces can be visualized in three dimensions. This is achieved through a process called dimensionality reduction \cite{wanright}. Another approach is to divide a higher dimensional space into multiple three-dimensional spaces. 
\vspace{0pt}
\section{conclusions}
In this work, we introduced a new probability distribution with a remarkably simple but diverse form of PDF. Then, we derived its statistical measures, all of which are in closed forms. Additionally, we provided asymptotic cases, one of which approaches Gaussian distribution. Furthermore, we made another new discovery by finding another probability distribution, which includes the Gamma distribution as a special case. We highlighted diverse applications of the proposed probability distributions in the fields of pure mathematics, statistics and learning, stochastic geometry, signal processing, and communication systems and networks.

\end{document}